\documentstyle[12pt,aaspp4]{article}

\def\msolar{ {M_{\odot}} }

\slugcomment{Accepted at The Astrophysical Journal}
\lefthead{Briggs}
\righthead{LSB Galaxies}

\begin{document}

\title{Estimation of the Space Density of Low Surface Brightness Galaxies}
\author{ F. H. Briggs}
\affil{Kapteyn Astronomical Institute, Postbus 800, 9700 AV Groningen, 
The Netherlands}

\slugcomment{Submitted to The Astrophysical Journal}
\lefthead{Briggs}
\righthead{LSB Galaxies}

\begin{abstract}

The space density of low surface brightness and
tiny gas-rich dwarf galaxies are estimated for two recent catalogs:
The Arecibo Survey of Northern  Dwarf and Low Surface Brightness
Galaxies (Schneider, Thuan, Magri \& Wadiak 1990) and
The Catalog of Low Surface Brightness Galaxy, List II (Schombert,
Bothun, Schneider \& McGaugh 1992).  
The goals are (1) to evaluate the additions
to the completeness of the Fisher and Tully
(1981a) 10 Mpc Sample and (2) to estimate whether the density of
galaxies contained in the new catalogs
adds a significant amount of neutral gas mass to the
the inventory of HI already identified in the nearby, present-epoch universe.
Although tiny dwarf galaxies
($M_{HI} {\lower .7ex\hbox{$<$}\atop \raise .2ex\hbox{$\sim$}}10^7\msolar$)
may be the most abundant type of extragalactic stellar system in the
nearby Universe,   
if the   new catalogs are representative,  
the LSB and dwarf populations they contain
make only a small addition 
($\lower .7ex\hbox{$<$}\atop \raise .2ex\hbox{$\sim$}$10\%) 
to the total HI content of the local Universe and 
probably constitute  even smaller fractions
of its luminous and dynamical mass.

\end{abstract}


\section{Introduction}

The goal of this paper is to decide whether there is already
evidence that substantial masses of neutral gas have been 
missed by belonging to such dim galaxies that they have 
escaped inclusion in the optically selected samples on which HI
studies historically have been based. New extensive catalogs of HI observations
of dwarf and low surface brightness (LSB)
galaxies exist in the literature. Do these new
objects make a significant addition to the integral HI content
of the Universe?
The question is of central importance to studies of galaxy evolution
that monitor neutral gas content as a function of redshift,
such as damped Lyman~$\alpha$ absorption-lines in the spectra of
quasars (Wolfe et al 1986, Wolfe 1988, Lanzetta  1993, Rao \& Briggs 1993,
Rao et al 1995).

Since the identification of LSB galaxies is severely hampered by 
the brightness of the night sky (Arp 1965, Disney 1976, 
Davies 1993), catalogs that rely on optical selection
to specify the numbers of these objects
are only now being constructed, through the use of new plate material
(Binggeli, Sandage \& Tammann 1985, Schombert \& Bothun 1988) 
and plate scanning machines
(Irwin et al 1990, Miller \& MacGillivray 1994, Impey et al 1994).
A substantial majority of the LSB systems is rich in HI
(cf. Longmore et al 1982, Schombert et al 1992), implying that
large unbiased radio surveys in the 21cm line may ultimately
be the most effective means of constraining their space density
(Zwaan et al 1997, Schneider 1997).

For the purposes
of this paper, the definition of ``LSB galaxies'' uses the
criteria that motivated Schombert and Bothun (1992) to include the galaxies in 
their catalog, namely that the objects are visible on 
the new plate material from the 
POSS II survey plates (Reid et al 1991) with sufficient angular size that
they would have been included in an angular-diameter limited sample such as the
UGC (Nilson 1973) if they had been of higher surface brightness.  The
issue is clouded by the fact that some galaxies have
low surface brightness disk components, which causes them to satisfy Schombert
and Bothuns' definition, and also have a bright, nuclear region
that would have caused them to be included in catalogs of normal surface
brightness galaxies  if they were located nearby, where the nuclear region
would subtend a larger angular extent.  Thus, there is a danger that a
portion of the population of nearby
gas-rich dwarf  galaxies would be perceived as  LSB disk systems
when they are detected at higher redshifts.  Alternatively, the danger is
that distant objects in new catalogs of LSB galaxies are already well
represented in the historically important catalogs that have
been used to compute luminosity functions and the local luminosity
density. 

In this paper, the 21 cm line luminosity of optically
selected objects is used to parameterize the newly identified
population of LSB galaxies in a way that permits 
their space density to be estimated. The additional contribution of this
population to the  HI-mass Function of the local 
Universe 
can then be compared to the number of systems
that comprise the bulk of the entries in large galaxy
catalogs such as the CGCG (Zwicky et al 1961) and 
RC3 (de Vaucouleurs et al 1991).  Computation
of the 
integral HI content follows trivially from the HI-mass function and finds
immediate application in the interpretation of quasar absorption-line
statistics.  Due to the  HI-richness of the LSB population 
(Schombert et al 1992),
the HI-mass function also helps to appraise the contribution of the
LSBs to the galaxy population as a whole.

The task is simplified by the existence of large catalogs of
21cm line observations.
In this paper, the analysis that Briggs and Rao (1993) applied
to the Fisher-Tully Catalog
(FT, 1981a) is supplemented by (1) the Arecibo Survey of Northern Dwarf
and LSB Galaxies (Schneider et al 1990), which relies on the UGC
(Nilson 1973) diameter-limited survey to fill in the LSB gaps in
the FT Catalog, and (2) the Catalog of Low Surface Brightness Galaxies,
List II (Schombert et al 1992), which applies the UGC diameter 
limit of 1$'$ at the 26.0 mag/square arcsecond isophote of the POSS-II,
which is ${\sim}$0.8 mag/square arcsecond fainter than the UGC limiting
isophote.
Throughout this paper, the Arecibo Survey of Schneider et al (1990) 
will be referred to
as the ``DLSB Sample'' and the Catalog of Schombert et al (1992) 
will be called the ``P2LSB
Sample.''  Since these two HI surveys of optically-catalogued galaxies 
were conducted
at  similar sensitivity levels with the Arecibo Telescope, they
provide large samples, whose selection criteria
were designed in a way that complements 
the earlier FT catalog of HI observations
of predominantly bright galaxies.  The 
sensitivity of the Arecibo Telescope ($\sim$10 times that of 
the telescopes used in the
FT observations) is necessary to study the dwarf and LSB systems, whose
HI fluxes tend to be low due either to their smaller physical size or 
greater average distances than the higher surface
brightness objects that dominate the FT Catalog.

As an introductory overview, note that only 12 galaxies from the
DLSB and none of the P2LSB sample are located in the volume 
used to define the FT 10 Mpc Sample of 355 galaxies detected in HI,
once the overlap between the DLSB and FT catalogs is removed.
The 12 DLSB galaxies have small HI masses 
($M_{HI}\leq 10^{8.5}\msolar$). Despite the fact that the
DLSB is drawn from roughly half the solid angle of sky of the FT sample, and
the P2LSB is taken from $\sim$15\% of the FT solid angle, the clear
implication is that the addition of these new samples cannot add much to 
the HI content already identified in the population of galaxies described
by the FT sample.  

In merging the space densities computed for the samples separately, 
there is a danger of double-counting
galaxy populations. For example, 
although the solid angle, $\Omega$, of the P2LSB  
is much less than the FT, the sensitivity-limited depth, $d_c$, of
the P2LSB   is roughly four
times deeper than the FT depth.  Since the survey
volume is ${\propto}d^3\Omega$,  the identification of $N$ low
surface brightness galaxies in the FT Sample, implies that this same
population should have 
${\sim}64N{\times}0.15\approx 10N$ representatives in the P2LSB. 
The space density computed from the larger volume of the P2LSB  
must not be added to the space density of galaxies
determined from the FT Sample unless we are certain that we are 
adding a population that is not already represented.
 
As selection of LSB objects is pushed to smaller angular diameters, 
there is also a danger that, without redshift information, more
distant occurrences of familiar populations will be mistaken for 
a very high space density of small nearby objects. The surface
density, number of
objects per steradian, will rise by a factor of 8 when the angular diameter
limit is halved. Thus, the measurement of redshift is a crucial 
ingredient to the interpretation of LSB galaxy catalogs.

A description of the analysis of the catalogs fills the bulk of this
paper. Section 2 describes the galaxy samples
in more detail. Section 3 is an account of the steps in the analysis, 
leading to the construction of   HI-mass
functions for the two samples. Section 4 compares the derived
HI-mass functions with that of the more familiar ``high
surface brightness'' (HSB) population, and
Section 5 attends to the correction of the FT 10 Mpc Sample for  
incompleteness.
Bounds on the space density of LSB galaxies are derived in  Section
6 along with an estimate of an LSB HI-mass function for
the types of galaxies added by the Catalog of Schombert et al (1992).
The comparison between the total HI content contained in the different
nearby galaxy populations is made in Section 7.
While there are numerous uncertainties
in this estimation process, it is
unlikely to be in error by a factor of two and represents an attempt
to evaluate the contribution of the LSB population relative to the
high surface brightness galaxies. 
A Hubble constant $H_o = 100$ km~s$^{-1}$~Mpc$^{-1}$ has
been used throughout.

\section{Properties of the Samples}

This section is a review of the properties of the FT, DLSB and P2LSB
Samples. Together they form a complementary picture of the HI-rich
segment of the nearby galaxy population.

FT constructed their source list for observation in the 21cm line
in order to obtain a high degree of completeness for late-type
galaxies within 10 h$^{-1}$ Mpc and with angular diameter
$a \geq 2'$; the vast majority of the FT sample is listed in the
UGC.  The DLSB Sample contains the  remainder of the UGC dwarfs and
LSB galaxies that fall within the declination range 
$-2^{\circ}$ to $+38^{\circ}$
accessible to the Arecibo Telescope, permitting observation of 
their spectra for 
HI emission over the redshift range $-400$ to $+6500$ km~s$^{-1}$
(with a few observed to 12,000 km~s$^{-1}$). The success
rate for detection in the DLSB  
was very high (${\sim}$90\%), which is probably
a natural consequence of the morphological selection and the UGC
angular diameter constraint $a \geq 1'$, which practically
guarantees a detectable HI mass at Arecibo. The criteria of
Schombert et al. (1990) for selecting the P2LSB Sample from 
the new POSS II survey plates (Reid et al 1991)
specifically excluded galaxies that were already in the UGC, obtaining
a sample that adds to the UGC completeness at the nominal UGC isophotal
diameter, as well as creating a new sample with members
selected  at a more
sensitive isophote. 
The Arecibo observations of the P2LSB  routinely covered the
velocity range $-500$ to $+13,500$ km~s$^{-1}$ with some galaxies observed
to 32,000 km~s$^{-1}$.  Although Schombert et al (1992) are cautious about the
completeness of the P2LSB, they (Schombert \& Bothun 1988) 
use the added sensitivity (to the 26 
mag/square arcsecond isophote) of the P2LSB survey to evaluate the
incompleteness of the UGC at 11.5\%. Furthermore, Schneider et al (1992)
use this result to estimate the incompleteness of the DLSB 
Sample at 14\%; Schombert et al (1992) claim to have remedied the 
incompleteness of the DLSB Sample by augmenting it with 
the statistics from their deeper P2LSB Samples.  
Clearly, incompleteness and unknown selection effects
are likely sources of error. However, these samples formed the largest
available data base for these objects at the time of submission
of this paper,
and the three catalogs together form a highly complementary
description of the late-type galaxy population. 

The samples are summarized in Table 1, which lists the
number of galaxies in each catalog, the number that were observed
in the 21cm line, the number that were detected, and an estimate of
the solid angle of the surveys. The specifications cover both 
the full FT Catalog  and the restricted FT ``10 Mpc'' Sample.  
The DLSB Sample is divided into two
lists, depending on the confidence level that Schneider at al (1990)
had in the accuracy of the Arecibo measurement: 
DLSB-1 contains observations that are free from
observational complications, such as possible confusion with a nearby
companion or poor spectral baselines which might create a greater than normal
source of systematic error, and DLSB-2 lists galaxies for
which there {\it might} be a problem. 
Schombert et al (1992) divide their catalog into two lists, depending on
whether $a \geq 1'$ (here referred to as P2LSB-1) or 
$30'' \leq a <  1'$ (P2LSB-2);  the P2LSB-1 and P2LSB-2 together form the
P2LSB Sample discussed here, although some properties of the sub-samples
will  be examined separately.
For purposes of the present analysis, the solid
angle of sky over which the nominal degree of completeness of the DLSB  
is expected to apply was estimated to be ${\sim}$3.4 steradians
from Schneider et als' Figure 2, by excluding the portion of the sky where 
Galactic obscuration has clearly influenced the identification of the
galaxies.  The P2LSB   results from 
inspection of 97 POSS II fields. Each field covers 
$6.5^{\circ}{\times}6.5^{\circ}$, but edge effects cause 
incompleteness outside a $6^{\circ}{\times}6^{\circ}$ region.
The fields are spaced on 5$^{\circ}$ centers, so that 
overlap between neighboring fields can occur. However,
the sparse coverage of the sky provided by the partially
completed POSS II means that the overlap occurred for
only a few of the fields from which the P2LSB   was 
drawn (J. Schombert, private communication).
For the present application, the independent solid angle
per plate was estimated at $(25+36)/2 = 30.5$ square degrees,
producing a total of $\sim$0.9 steradians for the entire P2LSB
Catalog. 

The HSB and LSB populations are mixed in the
FT and DLSB Samples. One gauge of the HSB content of the FT Sample is
the fraction of the sample that is also in the CGCG, which is ${\sim}$99\%
HSB galaxies
(Bothun, private communication). Only 41 (13\%)
of the 302 FT 10~Mpc 
galaxies in the declination range of the CGCG are not in the
CGCG. Since all of the high-mass
FT galaxies, $M_{HI} > 10^9\msolar$, are in the CGCG, we infer that
there are
no large  LSB galaxies in the FT 10 Mpc Sample. 
(This inference could be in error if some members of the FT Catalog
have both bright central regions that cause them to be selected for
the CGCG and have extended LSB disks that would cause them to satisfy the
P2LSB criteria if they were more distant.)
At the low mass end,
we have determined that
the analytic expression, $f_L=0.34 \exp(-M_{HI}/3{\times}10^8\msolar)$, 
provides a rough description of
the fraction of FT galaxies not in the CGCG  as a function
of $M_{HI}$ over the range $10^7$ to $10^9\msolar$. 
The DLSB is a mix of classifications: dwarf, irregular, 
Sd-m, or later.  On the other hand, the P2LSB  was selected for
low surface brightness, but LSB galaxies already in the UGC were omitted
from the list.  These mixtures in the samples create ambiguities in the
analysis that prevent a clean calculation of the space density of a
pure LSB population; instead, the samples will be used in
this paper to place limits.

Distances to galaxies are determined according to the procedure
used by FT: for the majority of the galaxies, the distance,
$d = v_{GC}/H_o$, is computed 
using the 21 cm redshift velocity corrected to the reference frame of the
Galactic center, $v_{GC}$, (de Vaucouleurs et al 1991)
 and a Hubble constant $H_o = 100$ km~s$^{-1}$~Mpc$^{-1}$.
Galaxies within 6$^{\circ}$ of the core of the Virgo cluster are
assigned a distance of 10.7 Mpc, and any remaining 
galaxies with $v_{GC} \leq 100$ km~s$^{-1}$ are assumed
to be members of the Local Group at assigned 
distances of 1 Mpc.  As a result, eighteen galaxies from the DLSB-1, 
six galaxies from DLSB-2, and none 
from the P2LSB   were
assigned the Virgo distance.  Only one galaxy (from 
DLSB-2) was assigned membership of the Local Group.
Since  only a small fraction of the galaxies in
the DLSB and P2LSB   lie within 10 Mpc,  only small
increments to the integral HI mass are made by including the
DLSB and P2LSB in the total.

\section{Analysis of the Samples}

The strategy will be to first treat the two surveys as though they sampled
independent populations.  Then, tests will be applied to determine
which segments of the P2LSB and DLSB samples are really the same 
types of objects already in the FT sample.

The analysis follows the procedures of Briggs \& Rao (1993), which
they applied to the FT Catalog to derive an HI mass function that
compared well with HI-mass functions derived via other routes
(Briggs 1990, Rao \& Briggs 1993). The method requires that the
sensitivity depth of the survey be determined as a function of
neutral gas mass, $M_{HI}$. This depth, $d_c$, is determined from
the observations themselves but is consistent
with a theoretical understanding
of the sensitivities of  radio telescopes.
The HI mass $M_{HI}$ for each galaxy is determined from the measured
21cm line flux and the galaxy's Hubble distance.  
The detections are counted in bins
of HI mass, and, in order to
obtain the HI mass function,
the number in each $M_{HI}$ bin is divided by the
volume in which this HI mass would have been detected.  

A small simulation was performed to demonstrate how this 
analysis, which uses the radio detectability of an optically selected sample,
can be expected to produce a reasonable estimate of the objects'
space density with high efficiency.  
The simulation also illustrates how a $V/V_{max}$ test can be used to 
check the applicability of the sample.

A galaxy sample was created by randomly populating
a cubical volume with galaxies according to a Schechter function with
faint end slope $\alpha = 1.25$. An optically selected sample was drawn by
viewing the contents of the cube from one corner and keeping galaxies 
satisfying ``observational limits'' on either magnitude or diameter.
Since the samples being discussed in this paper are nominally diameter
limited, the sample illustrated in Figure 1 is diameter limited, with
$D \propto L^{0.4}$ according to the Holmberg relation (cf. Holmberg 1975,
Peterson, Strom and Strom 1979).  This imposes a distance cutoff $d_c$ for
the diameter-limited sample, $d_c\propto L^{0.4}$, which is slightly
different from the 
distance cutoff for a magnitude limited sample, $d_c\propto L^{0.5}$.
As discussed in more detail below, the distance cutoff for detectability in
the 21 cm line is $d_c \sim M_{HI}^{5/12}$;  coupling this
dependence on HI mass with the 
trend of increasing HI richness toward lower optical luminosities
(cf Fisher \& Tully 1975, Briggs \& Rao 1993), $M_{HI} \sim L^{0.9}$, 
one finds a trend for the
HI detection distance cutoff of $d_c \sim L^{0.375}$.  The similarity of
the diameter limited cutoff ($d_c\propto L^{0.4}$)
with the HI detectability distance suggests that
such an optical, diameter-limited
selection will provide a sample with relatively uniform
detection rates in the 21cm line across the full span of intrinsic luminosities.

The top panel of 
Figure 1 shows the distances of  3042 galaxies in a simulated, 
diameter-limited sample plotted as a function of HI mass. An additional random
scatter in HI to optical luminosity ratio was added, with $\sigma$ of
0.16 dex, to simulate observational errors and allow for variance in the 
properties of the galaxy population. 
The solid curves depict 3 choices of sensitivity-limited
HI detection distance: the lowest curve indicates a cutoff that detects
35.8\% of the optically selected sample, the middle curve detects 74.1\%,
and the top curve detects 99.8\%. The central panel illustrates the
$<V/V_{max}>$ ratio (c.f. Schmidt 1968)
as a function of $M_{HI}$ for each of the three
HI sensitivities.  The lower panel shows the HI mass functions computed from
the number of galaxies detected at each of the sensitivity levels in 
comparison with the analytic form of the HI mass function computed for
mass bins of 0.5 dex (cf Briggs \& Rao 1993). The plot shows that a
reasonable estimate for the HI mass function can be computed from the 
detection rate in an optically selected sample, provided the HI
observations impose a real sensitivity cutoff in the sample. In other words,
the HI observations must be able to detect only a fraction of the sample
in order to insure that the survey volume is bounded by the sensitivity
of the HI observation.
In these cases, it is clear that the two samples that faithfully reproduce the
mass function are also the ones that are successful in the $<V/V_{max}>$
test, producing ratios close to 0.5 across the entire range of
HI mass.  The simulated observation that detects nearly all the diameter-limited
sample produces $<V/V_{max}>$ of 0.2 to 0.3 across the mass range and fails
to recover the mass function by a factor of 1/2 to 1/3.

The
steps involved in evaluating the samples and constructing the mass
functions for the DLSB and P2LSB samples are illustrated 
in Figures 2 through 7. 
Figure 2 illustrates the survey depths for 
both Samples
by plotting the measured galaxy distances, $d$, as 
a function of $M_{HI}$.  
Smooth curves for the survey depth, $d_c$, are plotted in
Figure 2;
a complete discussion of the choice of functional forms 
is given by Briggs \& Rao (1993).  
A  reasonable description of the survey depth is 
$d_c \propto M_{HI}^{5/12}$, which differs from a simple 
inverse square law dependence ($d_c \propto M_{HI}^{1/2}$)
because galaxies of larger
mass rotate faster, spreading their HI flux over a larger velocity
width $\Delta V$ and lowering their detectability. (In brief,
the noise level, $\sigma$,
in a spectrum that has been optimally
smoothed to match the HI profile width is $\sigma \propto 1/\sqrt(\Delta V)$.
The minimum detectable HI mass is $M_{HI} \propto 5\sigma \Delta V d^2$.
A sort of HI Tully-Fisher relation has 
$\Delta V \propto M_{HI}^{1/3}\sin i$ for galaxies with inclination $i$
relative to the plane of the sky,
leading to the result that $d_c \propto M_{HI}^{5/12}\sin^{-1/4}i$.
In fact, enough
information exists in the FT catalog to make a first order correction
for inclination, which does indeed sharpen the detection boundary (Briggs
\& Rao 1993). On the other hand, since the 
inclination and size information is less well determined or not provided
for all the samples and the $\sin^{-1/4}i$ factor is substantially
different from unity for a only small fraction of a randomly oriented sample,
inclination corrections are not attempted for the analysis of the DLSB and 
P2LSB Samples.)
A still more pessimistic evaluation of the detectability of large masses
with $d_c \propto M_{HI}^{1/3}$, would
lead to flatter mass functions than those illustrated here.

Figure 2 also illustrates the boundaries to the sensitivity of the
FT observations and the 10 Mpc limit over which the FT Catalog
was intended to be
complete. Very few galaxies from either the DLSB or P2LSB   fall in
the volume where the FT observations could have detected them. The few DLSB
galaxies that fall both within 10 Mpc and within the $d_c$ detection-limited
distance cutoff lie close to the $d_c$ boundary in the mass range
$M_{HI}^{7.2}$ to $M_{HI}^{8.5} \msolar$. 

Figure 3 shows velocity width $V_{20}$ of the HI profiles,
measured at 20\% of maximum, as a function of $M_{HI}$.  
The FT Catalog lists $V_{20}$, while, for the DLSB Sample, 
Schneider et al (1990) list both $V_{20}$ and $V_{50}$, 
the profile width measured
at 50\% of maximum. Unfortunately, Schombert et al (1992)
tabulate only the $V_{50}$
widths for the P2LSB. For purposes of comparing the samples,
the $V_{20}$ widths for the P2LSB   were estimated by
noting that, in the  DLSB Sample,
the  $V_{50}$ widths are on average 26 km~s$^{-1}$
larger than $V_{20}$ (i.e., $V_{20} = V_{50} + V_c = V_{50} + 26$), 
although $V_c$ has scatter over the range
from $\sim$10 to $\sim$
50 km~s$^{-1}$. In order to illustrate the effect of this scatter,
the estimated $V_{20}$ for the P2LSB Sample
are plotted with error bars that indicate this range.
No adjustment for inclination has been 
applied to either sample.  

The curves plotted in Figure 3 were originally
derived for the FT sample (Briggs \& Rao 1993) and represent a sort
of ``HI Tully-Fisher Relation'' with 
$\Delta V = C_0 M_{HI}^{1/3}$; $C_0 = 0.35$ km~s$^{-1}\msolar^{-1/3}$
corresponds to edge-on galaxies, and $C_0 = 0.08$ km~s$^{-1}\msolar^{-1/3}$
would result from an inclination ${\sim}13^{\circ}$.
The progression from the higher surface brightness galaxies in the FT sample,
through the
DLSB   to the P2LSB   shows a trend of declining velocity
width that is especially apparent in the range  
$M_{HI} \approx 10^9\msolar$, where
there are very few galaxies in the P2LSB that lie 
close to the $C_0 = 0.35$ 
upper bound. This effect indicates that LSB galaxies may rotate
more slowly for a given $M_{HI}$, which would be consistent with the
detailed analysis by Sprayberry et al (1994). 
Alternatively, it could be an indication
that the selection of LSB galaxies biases the sample toward face-on 
galaxies and that the more highly inclined (edge-on)
LSB galaxies are preferentially included in
samples of higher surface brightness galaxies,
consistent with the statement by
Schombert et al (1992) that the survey selection techniques 
are biased toward
face-on systems.  Regardless of the cause, the 
narrower profiles of the P2LSB galaxies make them easier to detect
in HI and contribute to making the depth of the Arecibo
observations ${\sim}$10\%
greater for the P2LSB than for the DLSB Sample, as can be
seen in Figure 2. A second factor that contributes to making $d_c$ greater
for the P2LSB   than the DLSB is  the improvement in sensitivity
of the Arecibo instrumentation with time. The DLSB observations took place
in the period 1979 to 1986, and the P2LSB galaxies were observed after 1988.
A third consideration is that morphological selection and the
angular diameter limit of the UGC have conspired to place the vast majority
of the DLSB galaxies above the Arecibo detection limit; the apparent detection
boundary of the DLSB 
may be due to selection rather than due to telescope sensitivity.

The HI-mass functions were estimated 
following the procedure used previously for the FT catalog (Briggs
\& Rao 1993). The
volume is computed (from the analytic expression for the depth
and the estimated solid angle) within which a given $M_{HI}$
would have been detected.  This volume was divided into the
number of galaxies with $d<d_c$ that were actually detected within an 0.2 dex
wide bin centered on that $M_{HI}$. In order to plot a quantity that
does not depend on the width of the bin, the counts were scaled
upward by a factor of five to produce mass functions of ``number
per Mpc$^3$ per mass decade.''  Since few observations of
the DLSB   were made for redshifts greater than 8000~km~s$^{-1}$,
the depth of the DLSB   was truncated at 80 Mpc, and no galaxies
with greater redshift were included.  Similarly, the P2LSB   was
truncated at a distance of 150 Mpc.  

In the following section, the mass function computed from
the DLSB sample will not be used in a formal way for the estimation of 
density of low surface brightness objects, since it is actually a mixture
of morphological types and may not actually be a sensitivity limited
sample (as discussed above). Instead, the DLSB   will be used to add to
the completeness of the FT 10 Mpc Sample at the low mass end. On the
other hand, the P2LSB sample should have purer morphological selection, and
it makes sense to evaluate it using the $<V/V_{max}>$ test.  Figure 4
shows $<V/V_{max}>$ plotted in four bins spanning
four decades of $M_{HI}$. The test is shown for two formulations of
$d_c(M_{HI})$. The value  for $<V/V_{max}>$  is $0.483\pm .026$ for the
110 galaxies contained in the detection volume when $d_c \propto M_{HI}^{5/12}$,
and $0.489\pm 0.025$ for 129 galaxies when $d_c \propto M_{HI}^{1/2}$.

\section{Comparison of the Sample Mass Functions}

The HI-mass functions for the FT, DLSB and P2LSB Samples are 
compared in Figure 5.
In the sensitivity limited regime, the survey depth was computed from
$d_c = d_o (M_{HI}/{\rm M}_{\odot})^{5/12}$ Mpc 
with $d_o =$ 0.004, 0.01 and 0.011 for the
FT, DLSB and P2LSB, respectively.  Error bars, computed from
counting statistics, are typically smaller than the symbols, except 
near the low mass end where there are often only one or two galaxies
per bin (see also Figures 4 and 5).
Since Figures 2 and 3 indicated 
no apparent discrepancy in the behavior between the
two DLSB-1 and DLSB-2, the two sub-samples
have been combined in subsequent
calculations of the mass functions.
After finding no striking abnormalities in the two P2LSB sub-samples, 
they too were combined to form a net mass function.
As indicated in Table 1, not all members of
the P2LSB Sample were observed at Arecibo,
so additional correction factors ($199/155 =1.28$ for the P2LSB-1
and $141/101=1.40$ for P2LSB-2)
have been applied to the space densities
based on the assumption that the detection rates would be the same
for the unobserved galaxies as for those that were observed.
For both the DLSB and P2LSB, 
a small correction was applied to compensate for the
overestimation of the space density of nearby galaxies, due to
enhancement by the Local Supercluster;  Felton (1977) computes
a factor of 2.3 overdensity in luminosity functions based on
on samples of nearby galaxies. Galaxies within
15 Mpc were therefore weighted by 1/2 in the counts used to produce the
HI-mass function; as can be seen in Figure 2, this affects only
a tiny fraction of the galaxies in these samples. 

The histogram for the FT 10 Mpc Sample in Figure 5 has been taken from
Briggs and Rao (1993) with a scaling factor of 1/2 to compensate for the
overdensity of the Local Supercluster.  There are no galaxies
in the 10 Mpc Sample with $M_{HI} > 10^{9.8}\msolar$, and the HI-mass function
plotted for this high mass range is a ``maximum envelope mass function''
derived from the full FT Catalog in the following manner:
the
space density was computed for each HI-mass bin as the survey
depth was incremented in steps of 5 Mpc.   As the cut-off distance 
was increased from 5 Mpc to $d_c$,
the peak space density that is obtained
for each HI-mass was retained as an
upper envelope to the mass function. At the end of the calculation,
the maximum density reached in each bin was adopted for plotting in Figure 5.
This portion of the FT
mass function above $10^{9.8}\msolar$ 
has higher uncertainty than the
lower mass range for two competing reasons: the nature of the method biases
the result toward being too large, while the incompleteness of
the FT Catalog at distances beyond 10 Mpc leads to 
underestimation.  

An analytic curve describing the HI-mass function, $\Theta(M_{HI}/M_{HI}^*)$
(Briggs 1990) is also plotted in Figure 5.  This function,
\begin{equation}
\Theta(M_{HI}/M_{HI}^*)
   = \Theta^* (M_{HI}/M_{HI}^*)^{-0.23}\exp[-(M_{HI}/M_{HI}^*)^{1.11}]
\end{equation} 
where  $\Theta^*$ is 0.026 Mpc$^{-3}$ per mass decade, 
and $M_{HI}^* = 10^{9.9}\msolar$, was derived from an expression for
the optical luminosity function and an approximate description of
the HI richness of spiral and irregular
galaxies. The inability of the analytic function to
match the FT mass function in the large mass regime may be an indication
of incompleteness in the FT survey or it may result from a genuine
failure of this simple analytic form to describe the true galaxy population.
The analytic curve predicts that 9 galaxies with $M_{HI}>10^{9.8}\msolar$
should fall within the volume of the FT 10 Mpc Catalog.
An improved fit is obtained by adopting $\Theta^* = 0.030$ Mpc$^{-3}$ per mass decade and lowering the mass cut-off slightly to $M_{HI}^* = 10^{9.7}\msolar$.
The large mass regime will eventually be 
constrained by unbiased radio surveys of large volumes
in  the 21cm line (cf. Sorar 1994, Spitzak 1996, Zwaan et al 1997).

\section{Completing the FT 10 Mpc Sample with the DLSB}

The 12 galaxies from the DLSB Sample that fall
in the volume used for the FT 10 Mpc Sample make a small contribution
to the HI-mass function at the low mass end. 
Figure 6 shows the incremental HI-mass function from these 12 galaxies;
the same plot has the HI-mass function derived for the entire DLSB Sample.
The heavy solid histogram of Figure 5 shows the modified mass function
for the FT 10 Mpc Sample, after the incremental mass function has been added.

\section{Bounds on the Space Density of LSB Galaxies}

The HI-mass functions illustrated in Figure 5 convey the relative
space densities of the objects in the catalogs. The difficulty lies
in bounding the importance of the low surface brightness portion of
the population, which contributes to all three samples. In this section,
an approximate mass function for LSB galaxies is derived, after 
consideration of the limits provided by the Samples.

\subsection{Upper Bound}

An analytic curve (dashed) is drawn in Figure 5 to represent an upper
limit to the space  density of the LSB population that is already
contained in the FT sample.  At the low mass end, the curve results from
the fraction of the FT galaxies that
are missing from the CGCG, 
$f_L(M_{HI})\Theta(M_{HI}/M_{HI}^*)$. 
The empirically determined cutoff occurs at the same location and with
the same shape as
that observed for the HI-mass function of
irregular galaxies (Rao \& Briggs 1993), which was derived from the 
optical luminosity function for irregulars from Tammann (1986); clearly, the
similarity occurs because both curves are related to the same 
population. For the dashed curve plotted in Figure 5,
the cutoff imposed by $f_L$ is
halted at a density of $1/2000 = 5{\times}10^{-4}$Mpc$^{-3}$, since the
volume probed by the FT 10 Mpc Sample is ${\sim}$2000 Mpc$^{3}$, and
nothing can be said about galaxies that are so rare that none are 
expected in this volume. In fact, the space densities 
inferred from the P2LSB Samples are sufficiently low that almost no galaxies
of the type contained in the P2LSB  in the
mass range greater than $10^9\msolar$
are expected in the FT volume; at lower masses, both the P2LSB
and DLSB populations are compatible with the limit defined by $f_L$.
Only two galaxies of the variety represented by the DLSB mass function
are expected in the FT 10 Mpc volume for $M_{HI}>10^9\msolar$, and these could
be either dwarfs or LSB galaxies according to the mix in the DLSB Sample.

Given that the above discussion  limits the large mass LSB galaxies to
be rare, the DLSB Sample mass function itself can be used to form
a conservative upper bound. The bound is conservative
since the sample is known to be diluted with higher
surface brightness objects.  At the low mass end, the DLSB mass function
lies along the $f_L\Theta$ curve (Figures 5 and 6), 
implying that, for small HI-masses,
the DLSB Sample may be a good representation of the objects that are
missing from the CGCG. Furthermore, there is only one bin of the 
HI-mass function
for the 12 DLSB galaxies falling in the FT 10 Mpc volume
(heavy line with error bars in Figure 6)
that lies significantly 
above the $f_L\Theta$ curve, suggesting that even if all 12 were
LSB galaxies, the curve forms an adequate bound on their space density.

\subsection{Lower bound}

The P2LSB Sample is the only one of the three Samples that has been 
morphologically selected to be purely low surface brightness galaxies.
Therefore, the space densities defined by the HI-mass function (Figures 5 
and 7) for the P2LSB form
a hard lower limit.  Throughout the low mass regime, Figure 7 shows
a mass function derived for the P2LSB that does not exceed the
dashed curve, which represents the fraction of FT sample missing from
the CGCG.

\subsection{An analytic low surface brightness galaxy HI-mass function}

An estimate for the HI-mass function of the low surface brightness
population contained in the
P2LSB, DLSB and FT Samples is 
\begin{equation}
\Phi_{LSB}(M_{HI}/M_{HI}^{\dagger}) = \Phi^{\dagger}_{LSB} 
(M_{HI}/M_{HI}^{\dagger})^{-0.25}\exp[-(M_{HI}/M_{HI}^{\dagger})^{0.4}] 
\end{equation}
with $\Phi^{\dagger}_{LSB} = 0.06$ Mpc$^{-3}$ per mass decade 
and $M_{HI}^{\dagger}=10^{7.9}\msolar$. 
The function is shown in Figure 7.  At the low mass end,
this curve includes the DLSB and $f_L\Theta$ populations. At the large
mass end, it traces the mass function defined by the P2LSB Sample.
In choosing this form of the curve, we are assuming that the bulk of the
low mass galaxies in the DLSB and the FT galaxies not listed in the CGCG
are genuine low surface brightness galaxies. At the high mass end, we
are assuming that the DLSB sample is heavily diluted by non-LSB objects but
that the P2LSB Catalog forms a hard lower bound.

\section{LSB Contribution to the Total Density}

The agreement between the HI-mass function deduced here
and the HI-mass function derived earlier by Briggs (1990) from a
simple extrapolation of the optical luminosity function is remarkably good,
at least down to the mass range ${\sim}10^{7}\msolar$ where the
statistics become weak (Figure 5). There is no evidence of a deficit of
low mass objects in the range just below 10$^{8}\msolar$
suggested by the analysis of Weinberg et al (1991), and 
the gently rising slope of the low mass end indicates that
the tiny
LSB galaxies at the faint end of the mass function may be the most
numerous type of nearby galaxy.
The normalization and shape of the HI-mass functions deduced from
more detailed considerations of the
optical luminosity functions (Rao \& Briggs 1993)
agree well with the new computation 
in the mass ranges above 10$^8\msolar$.

The statistics for masses near $10^{6}\msolar$ are very poor; the
LSB galaxies could be the dominant depositories of these tiny
quantities of extragalactic HI, although their spatial densities
are constrained by surveys in the 21cm line to be no more than
a factor ${\sim}10$ above the extrapolation of the mass function
from larger masses (Sorar 1994, Zwaan et al 1997). On the other hand,
there is, as yet, no evidence that such low mass objects are
abundant enough to support even an extrapolation of the HI-mass 
function to  $M_{HI} = 10^{6}\msolar$, and observations of dwarf
galaxies in the
Virgo cluster may indicate a deficit of such low HI-mass objects 
in that dense environment (Hoffman, Lu \& Salpeter 1992).

Although the
small field galaxies may outnumber the large ones, 
they cannot contain much of the HI or luminous mass
in galaxies, due to vast importance of the HSB galaxies close to the
knee of the mass function.
The relative contributions of the FT, DLSB and P2LSB
samples to the integral HI content of the local Universe
are summarized in Figure 8.
The HSB galaxies with  $M_{HI}$ in the range  $10^{8.5}$ to
$10^{10}\msolar$ dominate the neutral gas content at
$z\approx 0$.  Extrapolating the trends for the DLSB and 
P2LSB   to the poorly determined, low-mass regime around
$10^{6}\msolar$ implies that  low-surface-brightness galaxies
still provide only a tiny fraction of the total HI.
The incompleteness of the FT catalog for galaxies with large
$M_{HI}$ is a concern that is being addressed by  on-going HI surveys
in the 21 cm line (Sorar 1994). 
Published observations only constrain the 
number density of ``Malin 1 type '' galaxies (Sprayberry et al 1993)
to contribute less than
${\sim}10^7\msolar$~Mpc$^{-3}$ (Briggs 1990), which is comparable to
the amount of HI in each 0.2 dex wide bin near the peak at 
$M_{HI}=10^{9.5}\msolar$.

The total HI-mass densities, obtained from the integrals over the
distributions for each of the three samples in Figure 8, 
are $7.28{\times}10^7h$, $4.5{\times}10^6h$ and
$2.5{\times}10^6h\msolar$~Mpc$^{-3}$ for the FT, DLSB and P2LSB  
respectively. The ``corrected'' FT 10 Mpc Sample contains 
$\rho_{HI} = 7.43{\times}10^7h\msolar$~Mpc$^{-3}$, while the
analytic expressions using Equation 1 yield $10.7{\times}10^7h$
and $7.7{\times}10^7h\msolar$~Mpc$^{-3}$ for the original and
revised choices of $\Theta^*$ and 
and $M_{HI}^*$, respectively.
As fractions of $\rho_{HI}$, the DLSB and P2LSB   contain 0.065 and
0.034 of the local HI, respectively. The integral of the analytic
function for the estimated LSB mass function, Equation 2, yields 
$4.9{\times}10^6h\msolar$~Mpc$^{-3} = 0.052\rho_{HI}$.
Since the DLSB and P2LSB mass functions
form upper and lower bounds to the mass function of the LSB populations
in these catalogs, LSB galaxies can contain between ${\sim}$3 and 7\%
of the HI in nearby galaxies.  

Some recognition should be made for the dominant uncertainties: (1)
The FT sample draws on galaxies locally in the region which is recognized
to have enhanced density.  The correction factor (Felton 1977) that was
used here has uncertainty. (2) The exact area of sky and degree of 
completeness of the P2LSB sample (Schombert et al ) are not well quantified.
(3) The extent to which variability in the assignment of morphological
classification might shuffle previously cataloged galaxies into the
LSB bin based on the presence of a extended LSB disk is not well
understood.  In spite of these uncertainties, the conclusion here is 
that the new searches at the level of POSS~II will not add 15\% more
HI (i.e. not double the 7\% upper limit discussed in the previous
paragraph) to the census that is based on complete samples of nearby
gas-rich galaxies.

These arguments imply that
the HSB galaxies dominate the HI content of the local Universe.
Since the LSB galaxies in these DLSB and P2LSB Samples as a class
are characterized by large ratios of HI to optical luminosity (Bothun
et al 1985), the LSB galaxies should be expected to provide
an even smaller fraction of the present epoch luminosity 
density.  Furthermore, the trend seen in Figure 3, indicating that
the LSB population rotates more slowly than the HSB galaxies of
similar HI-mass, implies that the LSB contribution to ``dynamical
mass'' ($\propto V_{rot}^2R$)
is also a smaller fraction of that already identified with 
HSB galaxies, since there is no evidence that LSB galaxies have
substantially larger gaseous extent, $R$, than HSB galaxies of
the same HI mass. This conclusion was reached by the detailed study 
by Longmore et al (1982). 
There is no evidence that LSB galaxies
contain more than a small fraction of the luminous baryonic or the dark
matter contents of the Universe.
Since the method used in this analysis
relies on measurements of HI content and redshifts for the
sample members, an LSB population that is free of neutral gas would
of course escape inclusion in this analysis entirely.

\section{Conclusions}

Analysis of recent catalogs containing substantial numbers of
low surface brightness galaxies shows
that the currently catalogued low surface brightness population does not
add significantly to the HI content
of the local Universe, although their tiniest members could still
be the most numerous
type of nearby galaxy. Since as a rule the LSB galaxies are relatively
rich in neutral gas compared to the high surface brightness population,
it is unlikely that these LSBs contribute significantly to the luminosity
density or dynamical mass density of the local universe.  
The effect of lowering the detection
threshold by $\sim$0.8 magnitudes from the old POSS to the POSS II, appears
to add less than $\sim$3\% to the integrated HI content of the
Universe. At this rate,
lowering the sensitivity to 28 magnitudes per square arcsecond would only
raise the integral HI content by another 
$\lower .7ex\hbox{$<$}\atop \raise .2ex\hbox{$\sim$}$8\%.  
In fact, if a significantly larger number of 
large gas-rich objects existed, it
would already have been identified in a variety of HI surveys (Fisher
\& Tully 1981b, Briggs 1990,
Weinberg et al 1991).

The bulk of the neutral gas content of the local Universe is bound into 
the larger, HSB galaxies in quantities typically $M_{HI} \sim 10^{8.5}$ to
$10^{10}\msolar$, consistent with the computation of the Rao and Briggs (1993),
which relates the present HI content to the significantly larger HI densities
that are measured by the analysis of
QSO absorption lines at high redshifts (cf Lanzetta 1993, Wolfe 1988,
Rao et al 1995).

In the time since these results in this paper 
were submitted for publication, several
related studies have been presented in literature.  Davies et al (1994)
found that giant LSB galaxies are at least an order of magnitude less
common that their normal-surface-brightness counterparts, a conclusion
consistent with that of Briggs (1990) as well as the mass functions in
Figure 7 of this paper. McGaugh et al (1995) concluded that LSB galaxies
comprise more than 1/2 the general galaxy population, while McGaugh (1996)
finds that LSB galaxies contribute 10-30\% of the total luminousity
density. Unlike the the analysis presented in this paper, these 
studies did not make use of redshifts, instead basing
the analysis on the surface density of objects on the sky.  It is
also likely that the P2LSB galaxies are extreme cases of LSB galaxies, so
that basing a volume density estimate on them alone will underestimate the
total number density that would be obtained if the extreme and intermediate
LSB populations were combined (cf. McGaugh 1996).  On the other hand, to
reiterate the principal conclusion of this paper, 
pushing down the low surface brightness limits has not produced a
big increment in the recognized mass content of the local Universe.

\acknowledgments

Greg Bothun and Stacy McGaugh have
contributed  enlightening discussions to the
development of this analysis; their comments are greatly
appreciated.
The author is also grateful to James Schombert and Otto Richter for 
providing computer
files containing the Low Surface Brightness Galaxy Catalog and the
Huchtmeier-Richter Catalog of HI Observations of Galaxies, respectively, 
and
to Sheflynn Sherer for assistance with data entry and compilation.
This research has made use of the NASA/IPAC Extragalactic Database (NED),
which is operated by the Jet Propulsion Laboratory, Caltech, under contract
with the National Aeronautics And Space Administration.
This work has been supported by NSF Grant AST 91-19930 and 
NSF Grant AST 88-2222.

\newpage

\begin{figure}
\plotone{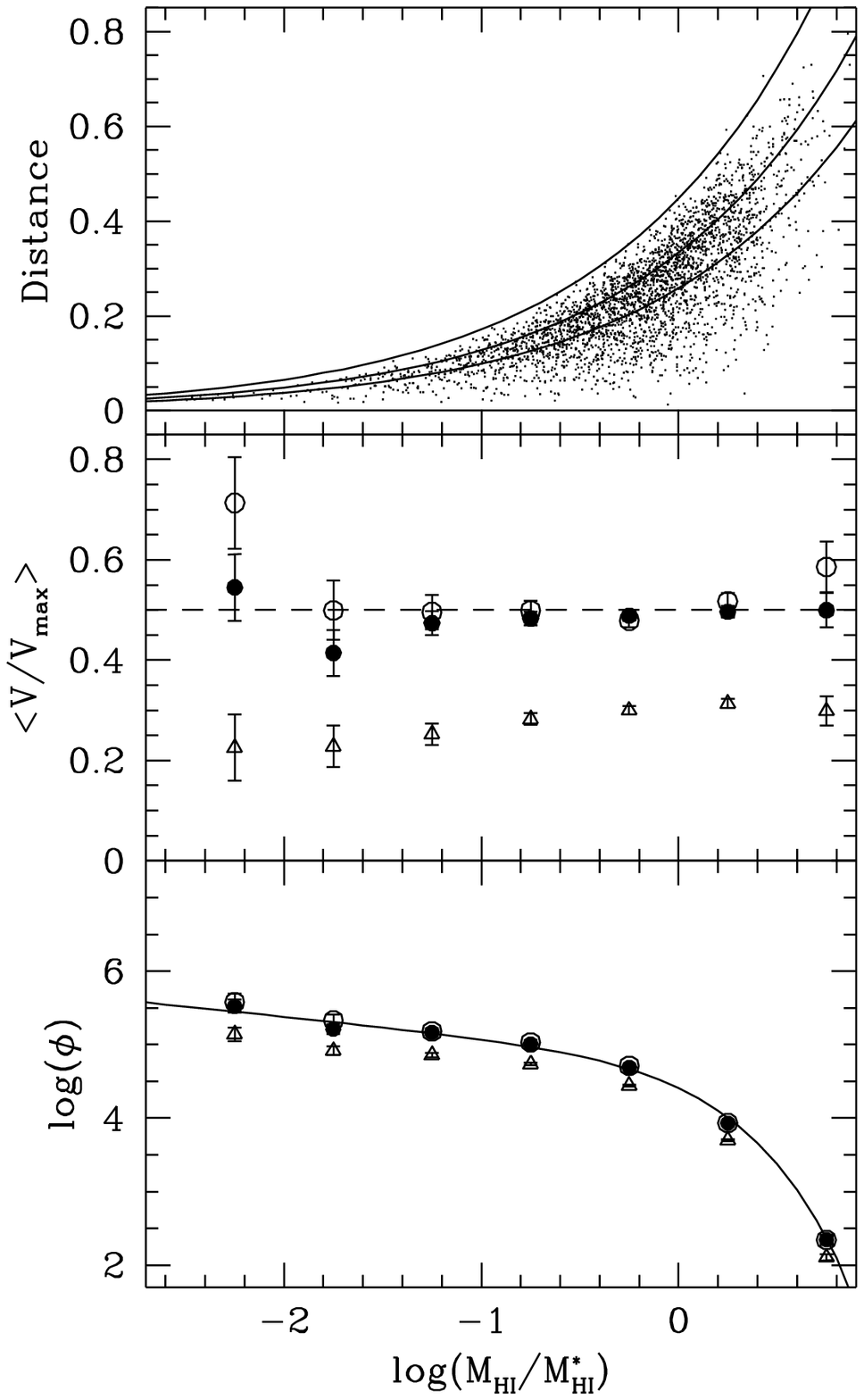}
\caption{
Simulation of the method for estimating the HI Mass function. {\it Top
Panel}: Distances of members of a diameter-limited sample of 3042 galaxies
plotted as a function of HI-mass. The curves represent sensitivity limited
cut-off distance, as a function of HI-mass, imposed by the radio observations
in the 21cm line. The simulated observational detection limits include
99.8\%, 74.1\% and 35.8\% of the 3042 galaxies, from upper to lower
curves, respectively.  
{\it Center Panel}: Results of $<V/V_{max}>$ test for the three simulated
detection sensitivities. Open circles for the 99.8\% cut, 
filled circles for 74.1\% and triangles for 35.8\%. 
{\it Bottom Panel} Points (as labeled in the center panel) indicate the
mass function derived from the simulated data for the three detection
sensitivities.  The solid curve is the HI-mass function expected for
the parameters used to generate the simulated sample.
}
\end{figure}

\begin{figure}
\plotone{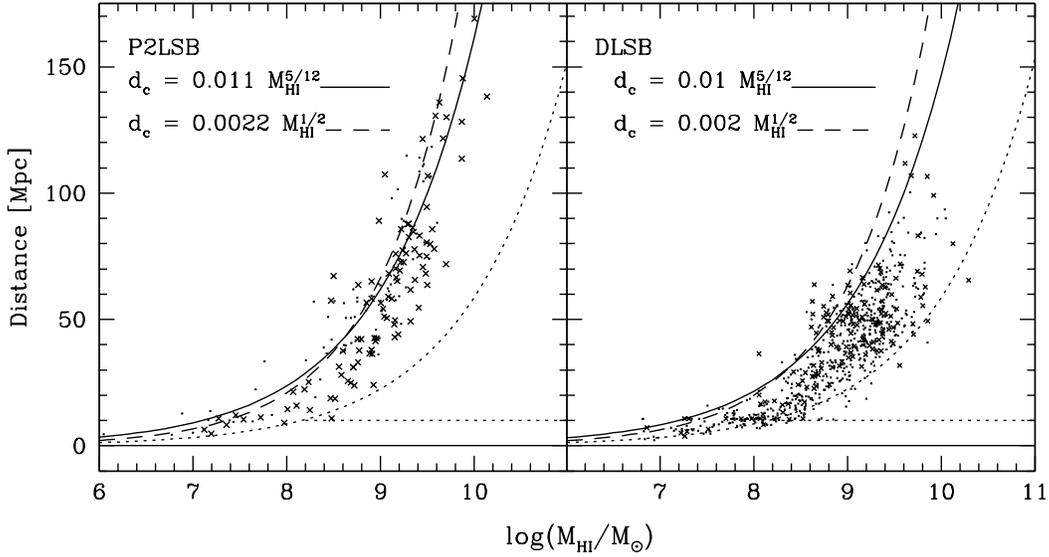}
\caption{
Distance as a function of HI mass for the P2LSB Sample {\it left} and
the DLSB Sample {\it right}. 
The solid curve indicates the
expected sensitivity limit  with
optimal spectral smoothing ($d_c \propto M_{HI}^{5/12}$) to
match the trend of increasing rotation velocity with increasing
HI mass. Also shown is 
$d_c \propto M_{HI}^{1/2}$ (dashed curve) that would be expected if galaxies
of all masses had the same rotation speed. 
For the DLSB, dots represent
galaxies from DLSB-1 and crosses stand for DLSB-2. For the P2LSB Sample,
dots represent galaxies with diameters
$30'' < a < 1'$ and crosses represent galaxies with $a > 1'$. The
dotted curves mark the sensitivity bound of the FT Catalog 
($d_c \propto M_{HI}^{5/12}$) and the 10 Mpc depth within which the
FT Catalog was designed to be complete.
}
\end{figure}

\begin{figure}
\plotone{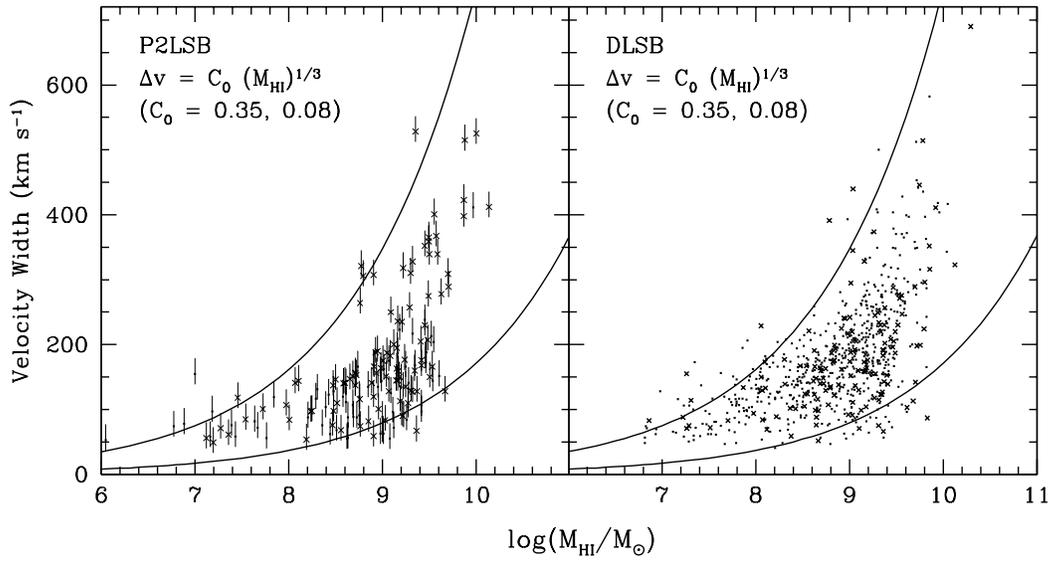}
\caption{
 Velocity width, $V_{20}$
as a function of HI mass  for the
P2LSB Sample, {\it left} and the DLSB Sample {\it right}.
No correction
for inclination has been applied to convert the data from
velocity spread to circular rotation velocity $V_{rot}$. The upper solid curve
forms an estimate of the dependence of $V_{rot} \approx 
V_{20}/2 \approx 0.18 M_{HI}^{1/3}$km~s$^{-1}$, taken from
the analysis (Briggs \& Rao 1993) for the Fisher-Tully Catalog (1981a). 
Error bars in the left panel indicate the
uncertainty in adjusting the P2LSB line widths to a system using 20\% of
maximum (see text). Dots and crosses are assigned as in Figure 2.
}
\end{figure}

\begin{figure}
\plotone{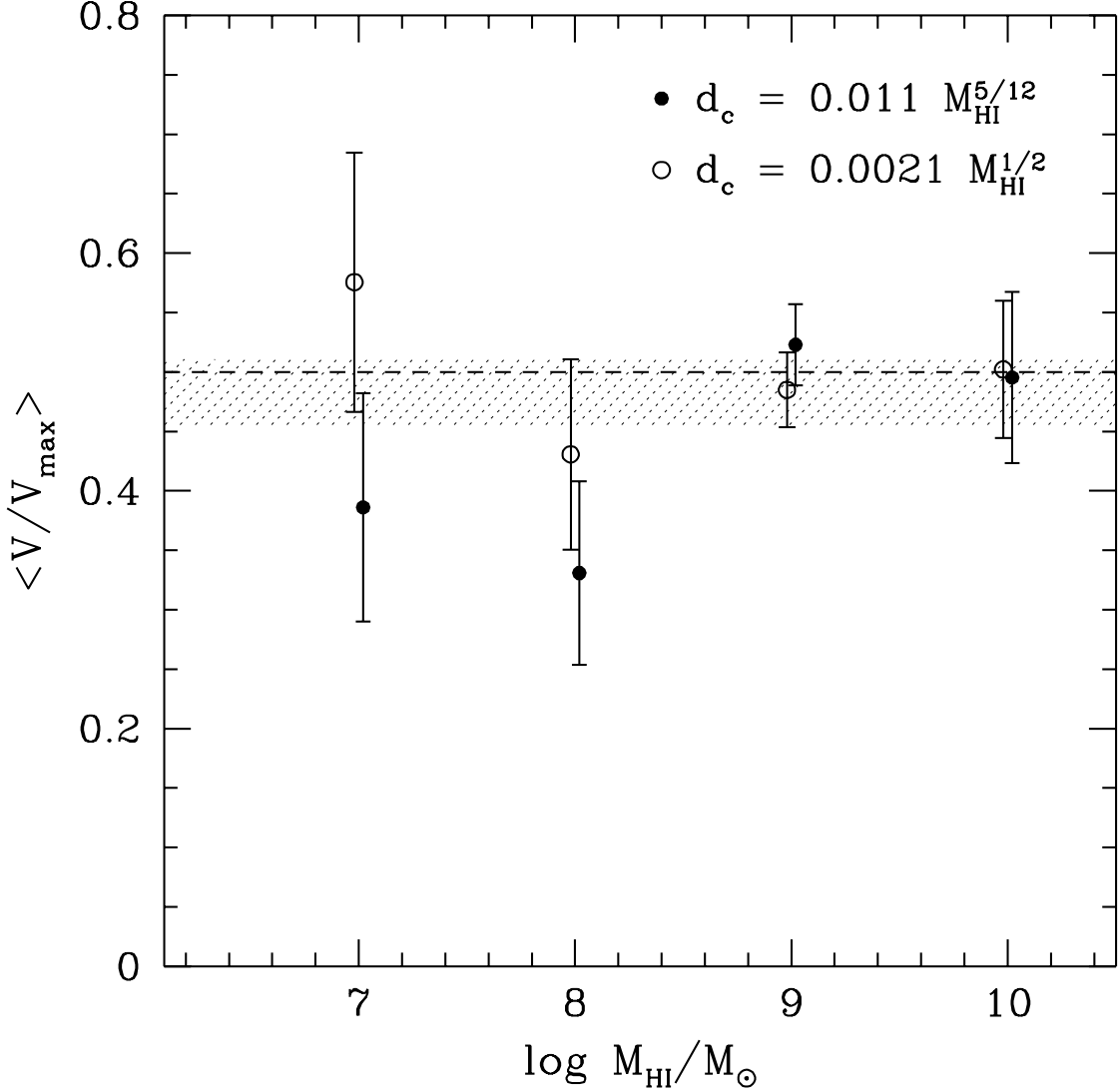}
\caption{
Results of the $<V/V_{max}>$ test for the 4 decades of HI mass sampled by the
P2LSB Sample.  Filled symbols represent a sensitivity limited distance cutoff
$d_c\propto M_{HI}^{5/12}$; open circles are drawn for 
$d_c\propto M_{HI}^{1/2}$.  The two cases have nearly the 
same $<V/V_{max}>$ (indicated by the shaded band)
for the entire sample analyzed together.
}
\end{figure}

\begin{figure}
\plotone{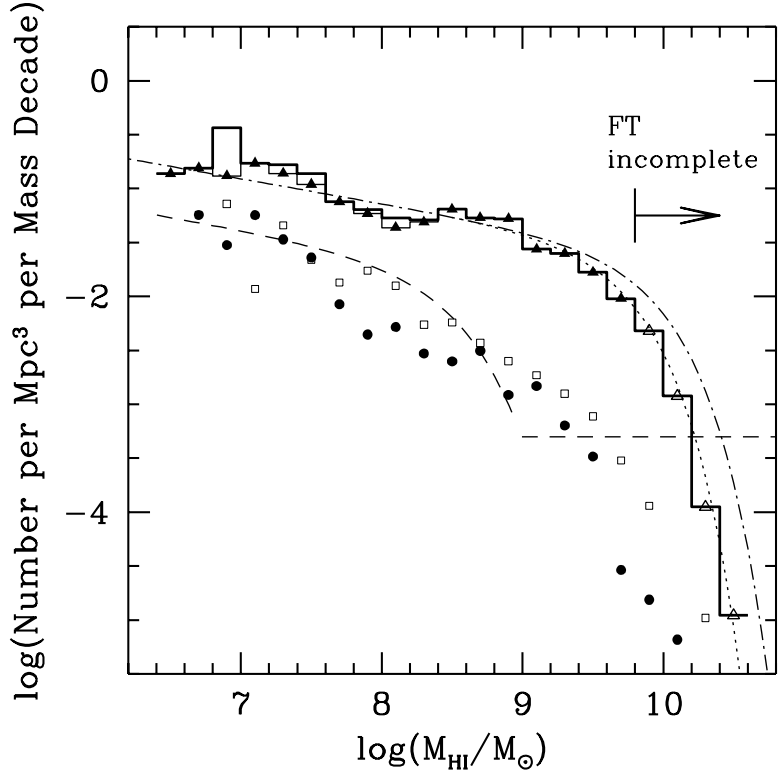}
\caption{
The total HI mass function and the mass functions of the FT, DLSB and P2LSB  
Samples. The total HI mass function, obtained by correcting
the FT 10 Mpc Sample, is drawn as a heavy, solid
histogram. The P2LSB  mass function is marked by filled dots;
the DLSB function  is marked by open squares; the FT 
function is marked by  triangles. The dot-dash curve results from Equation
1 with $\Theta^* = 0.026$ and $M_{HI}^* = 10^{9.9}$; the dotted curve
adopts $\Theta^* = 0.030$ and $M_{HI}^* = 10^{9.7}$. As described in the
text, the dashed curve is an upper bound on the LSB galaxy
component of the FT 10 Mpc Sample.
}
\end{figure}

\begin{figure}
\plotone{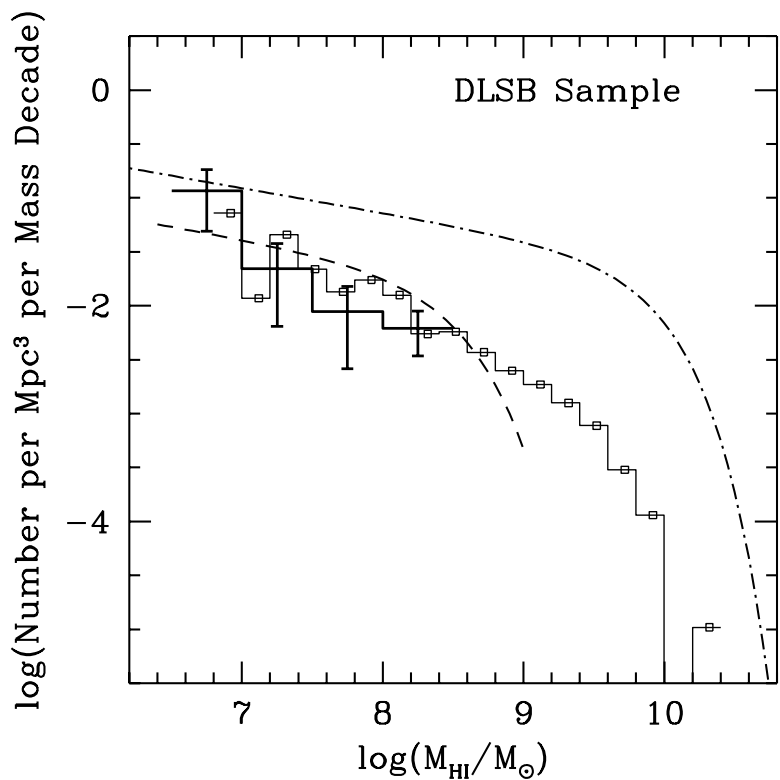}
\caption{
HI mass function determined for the DLSB Sample (unfilled squares)
The heavy solid histogram indicates the density computed for the
12 DLSB galaxies that fall in the FT 10 Mpc survey volume (bounded in depth
by either the FT sensitivity or 10 Mpc, whichever is less). 
Error bars based on
counting statistics are shown for the heavy histogram, which has been
binned in half-decades due to the small number of galaxies.  The dot-dash 
and dashed curves are those described in Figure 5.
}
\end{figure}

\begin{figure}
\plotone{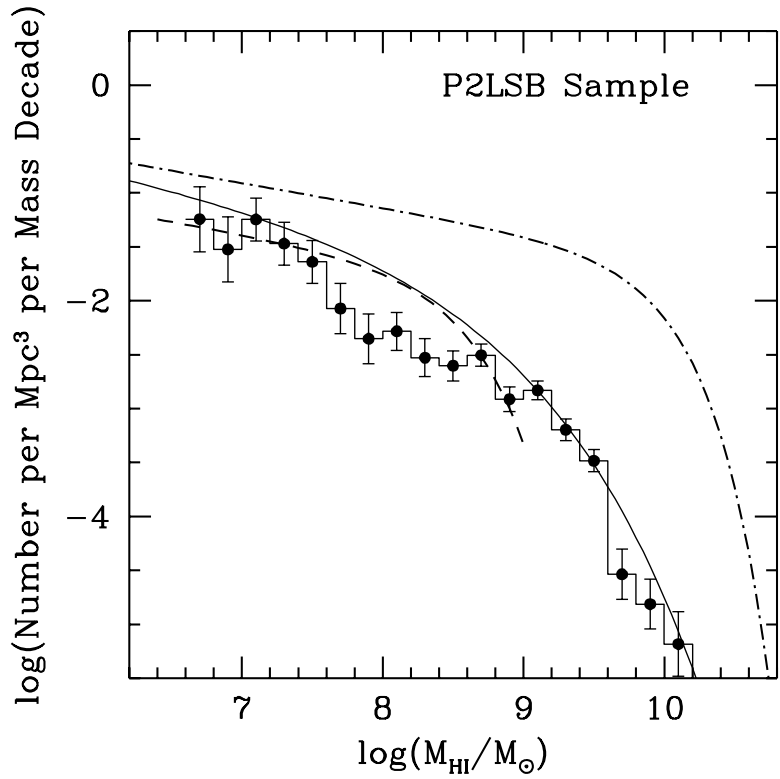}
\caption{
HI mass function determined for the P2LSB Sample, indicated
by heavy histogram and filled dots. Error bars are based on
counting statistics. The smooth solid curve represents the estimated
HI-mass function for LSB objects from Equation 2.
The dot-dash 
and dashed curves are those described in Figure 5.
}
\end{figure}

\begin{figure}
\plotone{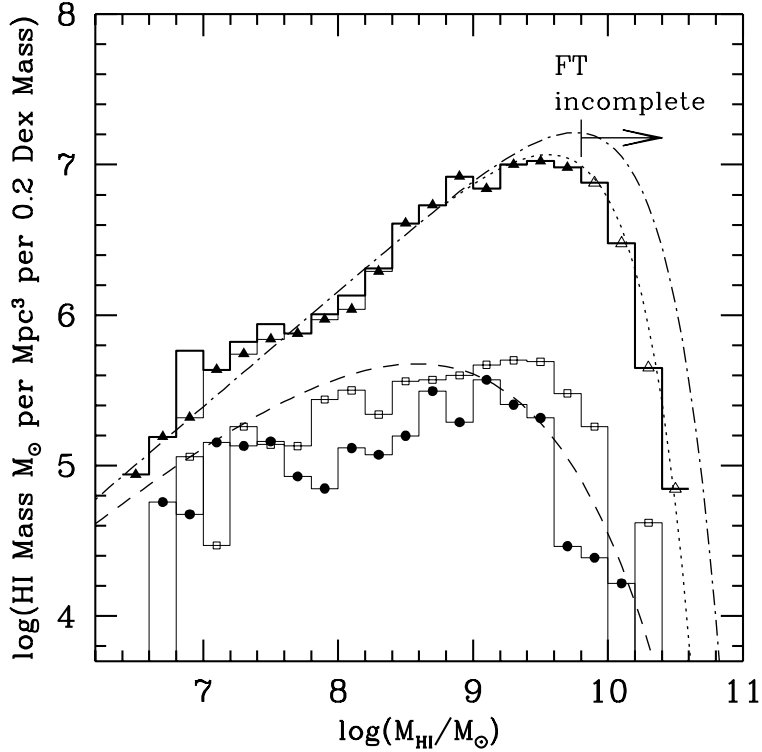}
\caption{
The integral HI mass density contained in catalogued populations of
galaxies as a function of the HI mass of the host galaxy. The
``corrected '' FT Sample is represented
as a heavy, solid histogram.
The P2LSB   Sample is marked by filled dots;
the DLSB Sample is marked by open squares; the original FT 10~Mpc
Sample is marked by  triangles. Smooth curves are drawn to indicate the
contents of the
analytical models: {\it dot-dash} integral HI-mass function  
($\Theta^* = 0.026$ 
and $M_{HI}^* = 10^{9.9}$),  {\it dots} integral HI-mass function
($\Theta^* = 0.030$ and $M_{HI}^* = 10^{9.7}$), and {\it dash} estimated
LSB HI-mass function.
}
\end{figure}

\newpage

\begin{deluxetable}{lrrrr}
\tablewidth{425.0pt}
\tablenum{1}
\tablecaption{The HI Observations}
\tablehead{
\colhead{Sample} & \colhead{Total Number} &
\colhead{Number} & \colhead{Number}    &
\colhead{Est. Solid } 
\nl
\colhead{Name} & \colhead{of Galaxies} &
\colhead{Observed} & \colhead{Detected}    &
\colhead{ Angle [Sterad]}
}

\startdata
FT           & 1720  & 1720 & 1153  & ${\sim}2\pi$   \nl
~ ~FT 10 Mpc\tablenotemark{a} & 357 & 357 & 343 & \nl
           &         &      &      &              \nl
DLSB\tablenotemark{b}& 762  & 762  & 709  & $3.4$ \nl
~ ~DLSB-1     & 574  &      & 574 &               \nl
~ ~DLSB-2     & 135  &      & 135 &               \nl
           &         &      &      &              \nl
P2LSB-1\tablenotemark{c} & 199  & 155  & 98   & $0.9$        \nl
P2LSB-2         & 141  & 101  & 72   &              \nl 
\enddata
\tablenotetext{a}{The FT 10 Mpc Sample is a subset of the
main FT Catalog}
\tablenotetext{b}{The detections in the  DLSB Survey
are divided into two subsamples: DLSB-1 and DLSB-2}
\tablenotetext{c}{The detections in the P2LSB Survey
are divided into two subsamples, with P2LSB-1 comprising galaxies
of angular diameter greater than $1'$ and P2LSB-2 being less than $1'$.}

\end{deluxetable}

\end{document}